# Accurate hadron spectroscopy on blocked configurations by tadpole-renormalized, clover-improved Wilson quark action

Artan Boriçi and Philippe de Forcrand [a] [*]

[a]Interdisciplinary Project Center for Supercomputing (IPS),
Swiss Federal Institute of Technology Zurich (ETH),
IPS, ETH - Zentrum, CH - 8092 Zürich, Switzerland

We compare the quenched hadron spectrum on blocked and unblocked lattices for the Wilson quark action, the clover action and the tadpole-improved clover action. The latter gives a spectrum markedly closer to the original one, even though the cutoff is $a^{-1} \sim 500$ Mev.

A great deal of work has been done both on perturbative lattice QCD analysis and numerical simulations. Yet, recently it has been shown that more attention should be paid to simultaneously using both tools to make progress in this field [1]. Dramatic gains can be achieved by better lattice discretization of continuum operators. The Wilson quark action, with its O(a) discretization error, is the foremost candidate for improvement.

We study such a systematic improvement here, testing it on the quenched hadron spectrum. Since we do not know what the 'correct answers' are for the hadron masses, we use blocked configurations and try to reproduce the unblocked spectrum. We use $8^3 \times 16$ SU(3) gauge configurations obtained by 2 blocking steps from a $32^3 \times 64$ lattice at $\beta = 6$ [2]. The blocking procedure preserves the infrared properties of the gauge field. On the other hand, the quark action is expected to show stronger lattice artifacts, which will be the object of two improvements: the O(a) correction [3] known as the 'clover' correction, and the 'tadpole' improvement, which accounts for large tadpole renormalization of the link variables [1]. This will increase the fill-in of the corresponding matrix and therefore the computer time needed for matrix-vector multiplication. This is more than offset by the possibility of going to a coarser lattice, where $ma$ is also larger. In our case, the gain from the volume alone is a factor 256.

For matrix inversion, we use the method of stabilized biconjugate gradients BiCGstab2 [4], which we show to be faster than conjugate gradient (CG) and conjugate residual (CR(1)).

## 1. MATRIX INVERSION

The standard linear solvers in QCD simulations have been CG and CR in its CR(1) version [5]. CG solves the normal equations so that the condition number is squared, resulting in slow convergence. CR(1) is slow compared to its full version and convergence is not guaranteed.

Alternatively, one can use the biconjugate gradient (BiCG) algorithm, which constructs two sequences of mutually orthogonal residual vectors. This is done by applying the nonsymmetric Lanczos method to the original matrix and to its adjoint. The method does not imply minimization of the residual norm $||r|| = ||b - Ax||$, therefore leading to a very irregular convergence history of $||r||$. The newer method of BiCGstab [7] stabilizes the residual polynomial of BiCG by applying to it another polynomial, i.e. $r_{BiCGstab} = S(A) r_{BiCG}$, where $S(A)$ is constructed in factors $(1 - \chi_1 A)(1 - \chi_2 A)\ldots$. At each step $k$, $\chi_k$ is chosen so as to minimize the $||r_k||$ of BiCGstab. BiCGstab2 does this minimization in the two-dimensional complex space of $\chi_{k-1}$ and $\chi_k$ every other step, producing faster convergence. In Figure 1 it is shown that BiCGstab2 always beats CR(1) and CG by a factor $\sim 1.3$ in the weak and strong coupling limits, respectively [6].

[*]This work is done under Grant No. 2100 - 037744.93 of the Swiss National Foundation for Scientific Research

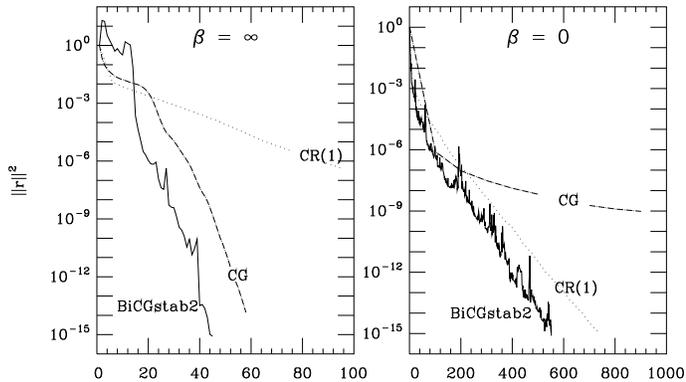

Figure 1. The convergence behaviour of different matrix inversion algorithms for the Wilson quark action on an $8^4$ lattice. $\kappa$ is adjusted so that $\lambda_{max}/\lambda_{min} = 200$.

Further improvement can be obtained by block methods, which solve a given linear system for $s$ different right hand sides. In this case the dimension of the space spanned by the Lanczos vectors will be expanded by a factor $s$ so that the number of iterations can be expected to decrease by the same factor. We have implemented block BiCG and see clear improvement on a small lattice.

## 2. RESULTS OF SPECTROSCOPY

We have calculated quark propagators and hadron masses for 55 blocked configurations. Mass errors are estimated by jack-knife. Using the Wilson action as the quark action, the $\rho$ mass was extrapolated through a quadratic fit to 0.19(3) in units of the unblocked lattice $(a')$. (To rescale the lattice spacing we have matched the string tension $(\sigma a^2 = 0.93)$ to its value on a $32^4$ lattice at $\beta = 6$ from Ref. [9]). This value is far from the value 0.34(1) extrapolated from the data of [8] on the unblocked lattice using the Wilson quark action. The discrepancy comes from discretization errors on our coarser grid. Therefore, we tried to improve the Wilson action systematically.

The 'clover' O(a) correction [3] to the above action gave us $m_\rho = 0.25(1)$, which shows some improvement.

Recently, it has been shown how important tadpole renormalizations are as one tries to connect perturbatively the lattice approach to the continuum theory [1]. This prescription consists simply of dividing each link by $u_0 = \langle \frac{1}{3} Tr U_{plaq} \rangle^{1/4}$. This amounts to a trivial rescaling of the hopping parameter $\kappa \to \kappa u_0$ in the Wilson action, leaving $m_\rho$ unchanged (see Table 1).

We then tried to combine clover and tadpole improvement. In this case the extrapolated $\rho$ mass is 0.30(2), markedly closer to the value 0.34(1) taken as reference. In Fig.2 $\rho$ masses are compared for the different quark actions, showing the important effect of tadpole renormalization. Complete spectroscopy results are presented in Fig. 3 for the doubly improved action.

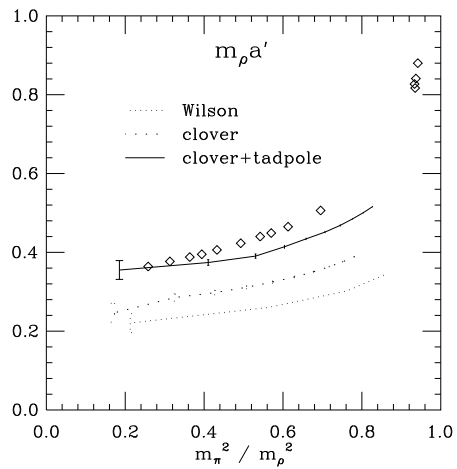

Figure 2. The $\rho$ mass for different quark actions on $8^3 \times 16$ blocked configurations. Diamonds are the results from Ref. [8].

The similarity to the unblocked results of [8] is remarkable, given that we are working at a cutoff $a^{-1} \sim 500$ Mev (equivalent to $\beta \approx 5.1$). Note that we can reach smaller quark masses than in [8] at a small fraction of the cost, including the regime $m_\pi/m_\rho < 1/2$.

Another indirect measure of improvement is the approach of the critical hopping parameter $\kappa_c$ to its free value $1/8$. The values of $\kappa_c$ listed in Table 1 again show the benefits of the doubly improved action.

Improvement of the action is visible as well in the eigenvalue spectra. In Fig. 4 we show the eigenvalues of matrices corresponding to differ-ent formulations of quark actions. They are compared to the spectrum of free quarks on a $32^4$ lattice.

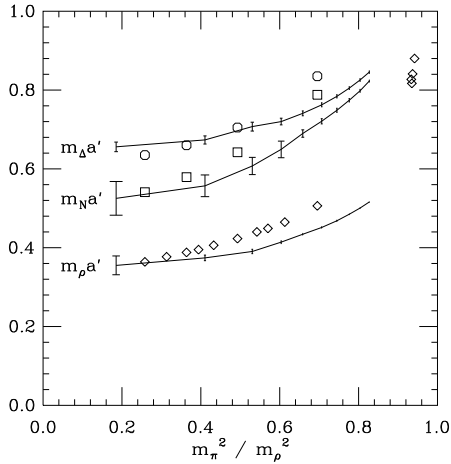

Figure 3. Hadron masses for clover-and-tadpole-improved Wilson quark action on $8^3 \times 16$ blocked configurations. Diamonds, squares and octagons stand for $\rho$, $N$, and $\Delta$ masses of Ref. [8].

| Quark action | $\rho$ mass | $\kappa_c$ |
|---|---|---|
| Wilson | 0.19(3) | 0.205 |
| tadpole | 0.19(3) | 0.163 |
| clover | 0.25(1) | 0.175 |
| clover+tadpole | 0.30(2) | 0.117 |

Table 1
Changes in $\kappa_c$ and $m_\rho$ for different quark actions.

The similarity of the low-lying spectrum of the doubly improved action to the free one is remarkable. This result suggests that the free Wilson matrix be used as a preconditioner for the linear solver. After fixing to Landau gauge, preliminary results show that the matrix inversion can be Fourier Accelerated at least by a factor of two, which will be much greater at higher $\beta$. The acceleration was also greater as $\kappa$ was increased, suggesting that small quark masses can be simulated efficiently.

## 2.1. Acknowledgements

We thank the QCD-TARO collaboration for generating our blocked configurations, and R. Gupta for communicating his data to us.

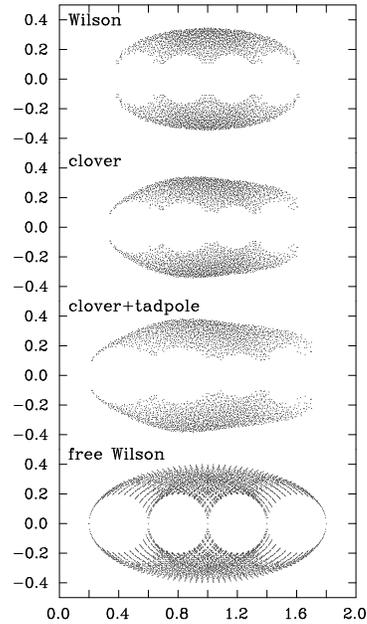

Figure 4. Comparison of spectra corresponding to different quark actions for $\kappa = 0.1$. The background field in the first three cases comes from a $4^4$ lattice at $\beta = 6$, and the last spectrum is that of free quarks on a $32^4$ lattice with periodic b.c.